\documentclass[10pt,a4paper]{article}

\usepackage{graphics}
\usepackage{graphicx}
\usepackage{dcolumn}
\usepackage{bm}
\usepackage{amsmath,amssymb}
\usepackage[affil-it]{authblk}

\newcommand{\de}{\partial}

\newcommand{\eq}[2]{\begin{equation} \label{#1} #2 \end{equation}}

\newcommand{\etal}{{\em et al.}}

\begin{document}

\title{\bf Unphysical metastability of the fundamental Raman soliton in the reduced nonlinear Schroedinger equation}
\author{Truong X. Tran$^{1,2,*}$ and Fabio Biancalana$^{3}$\\
\small $^{1}$Department of Physics, Le Quy Don University, 236 Hoang Quoc Viet str., 10000 Hanoi, Vietnam\\
 \small $^{2}$Max Planck Institute for the Science of Light, G\"{u}nther-Scharowsky str. 1, 91058 Erlangen, Germany\\
 \small $^{3}$School of Engineering and Physics, Heriot-Watt University, EH14 4AS, Edinburgh (UK)\\
\small  $^{*}$Corresponding author: f.biancalana@hw.ac.uk}
\date{\today}

\maketitle

\begin{abstract}
We demonstrate theoretically and numerically that the fundamental Raman soliton of the widely used nonlinear Schroedinger equation (NLSE) with a linear approximation of the Raman gain ({\em reduced} NLSE) is metastable. It can propagate for hundreds of dispersion lengths along the optical fibre before eventually disappearing due to a peculiar instability, leading to a collapse. The noise eigenfunction analysis agrees well with the results obtained via direct pulse propagation simulations. This instability is not present when modelling the Raman effect via a full convolution, and thus the reduced NLSE often leads to unphysical results, and should be avoided.
\end{abstract}


The dynamics of a short and intense pulse in optical fibres is determined by several physical effects. One major effect is the stimulated Raman scattering (SRS) which becomes particularly important for sub-picosecond pulses \cite{agrawalbook1,kivshar}. In the presence of the Raman effect combined with Kerr nonlinearity and in the absence of higher-order dispersion (HOD), Raman states with a single or double peak structure were numerically found by solving the {\em reduced} nonlinear Schroedinger equation (NLSE), where a linear approximation of the Raman gain is used [see Eqs. (\ref{NLSE}) - (\ref{Gagnon}) below], by Akhmediev \etal \cite{akhmediev}. In Ref. \cite{akhmediev}, the Raman two-peak states were also shown to be {\em metastable} in the sense that they can propagate for long distances before eventually disappearing due to internal collapse. A few years ago, in a series of experiments \cite{pod1,pod2}, short and intense pulses have been launched in highly nonlinear photonic crystal fibres (PCFs, see also \cite{russell}). The formation of long-lived, two-peak soliton states for specific 'magic' input pulse energies was observed. In these experiments, it was clear that this phenomenon is quite universal for a large variety of highly nonlinear optical fibres, especially in the deep anomalous group-velocity dispersion (GVD) regime far from zero-GVD points, where the impact of HOD is minimised. Such localised states always show peaks of differing amplitude with a well-defined peak power ratio $r$. Moreover, the temporal separation between the two peaks is also uniquely determined by $r$ and the Raman parameter $\tau_{R}$. A full theoretical understanding of the excitation of Raman multipeak states (with an arbitrary number of peaks) in solid-core PCFs was provided in Ref. \cite{tran1,tran2}. Based on a 'gravity-like' potential approach, in Ref. \cite{tran1} we derived simple equations for the 'magic' peak power ratio $r$ and the temporal separation between pulses forming these multipeak states. We also developed a model to calculate the magic input power of the input pulse around which the phenomenon can be observed. In Ref. \cite{tran1,tran2} we also numerically demonstrated that all Raman multipeak states are metastable, and they become more and more unstable when increasing the number of peaks. This instability occurs in {\em both} the reduced and the {\em full} NLSE, where in the latter case the Raman response function is used without approximations.

The physical metastability of Raman multipeak states has been thus proved unambiguously in our previous work \cite{tran1}. However, the stability of the {\em single} Raman peak state (further referred to as the fundamental Raman soliton, FRS in brief) has not yet been determined conclusively. In Ref. \cite{akhmediev}, the FRS was claimed to be stable via numerical simulations of the reduced NLSE for a propagation distance up to about 100 dispersion lengths. However, as we will show later, this propagation distance is not sufficient to prove that the FRS is stable in the reduced model. In Ref. \cite{Portuguese_papers}, the authors used arguments based on the Evans function, concluding again that the FRS of the reduced NLSE is stable. However, we show here that this is not the case.

Our analysis explains and confirms previous numerical simulations performed by Erkintalo \etal \cite{erkintalo1} and Solli \etal \cite{solli1}, who observed pulse collapse and energy non-conservation when using the reduced NLSE. In this latter paper, famous for the first experimental observation of rogue waves in optics, the authors remark that ``The rogue waves have a number of other intriguing properties warranting further study. For example, they propagate without
noticeable broadening for some time, but have a finite, seemingly unpredictable lifetime before they suddenly collapse owing to cumulative
effects of Raman scattering. [...] The decay parallels the unpredictable lifetimes of oceanic rogue waves". It is shown here that this collapse is totally unphysical and is a byproduct of the instabilities introduced by the linearized Raman gain, as hinted in \cite{erkintalo1} .

In this Letter we theoretically and numerically demonstrate that the FRS of the reduced NLSE is unstable: it can propagate for several hundreds of dispersion lengths along the fibre without significant changes in its profile, but eventually it will collapse due to a peculiar instability. However, long-distance simulations of the full NLSE conclusively show that the Raman soliton is perfectly stable in this physically more precise model. This underlines important physical inconsistencies of the reduced NLSE, which should therefore be carefully avoided for numerical modelling.

For wide enough pulses (temporal width $>100$ fs) the generalised NLSE governing the pulse propagation in silica optical fibres and solid-core PCFs can be simplified in dimensionless units in the laboratory frame ($z,t$) \cite{akhmediev}, by using a linear approximation of the Raman response function, giving rise to the {\em reduced} NLSE:
\eq{NLSE}{i\de_{z}\psi+\frac{1}{2}\de_{t}^{2}\psi+|\psi|^{2}\psi-\tau_{R}\psi\de_{t}|\psi|^{2}=0,} where $\psi(z,t)$ is the electric field envelope rescaled with the fundamental soliton power $P_{\rm S}\equiv |\beta_{2}|/(\gamma t_{0}^{2})$, $\beta_{2}$ is the GVD coefficient at the reference frequency, and $\gamma$ is the fibre nonlinear coefficient. Variables $z$ and $t$ are dimensionless space and time, respectively, rescaled with the second order dispersion length $L_{\rm D}\equiv t_{0}^{2}/|\beta_{2}|$ and with the input pulse duration $t_{0}$. The last term of Eq. (\ref{NLSE}), responsible for the Raman effect, produces a constant Raman soliton self-frequency shift (SSFS) in silica fibres \cite{mitschke,gordon}, $\tau_{R}\equiv T_{R}/t_{0}$ being a small parameter, where $T_{R}$ is the Raman response time, approximately equal to $2$ fs in silica. Solitons subject to the Raman effect shift continuously towards the red part of the spectrum due to SSFS, leading to a constant (negative) acceleration of the soliton in the time domain \cite{mitschke,gordon}.


In the reference frame of an intense accelerating soliton, one can operate the so-called Gagnon-B\'elanger gauge transformation $\psi(z,t)=f(x,\xi)\exp\left[iz\left(q-b^{2}z^{2}/3+bt\right)\right]$, where $x = z$, $\xi\equiv t-bz^{2}/2$, and $b=32\tau_{R}q^{2}/15$. This leads to the following equation for $f(x,\xi)$ \cite{belanger,akhmediev}:
\eq{Gagnon}{i\de_{x}f + \frac{1}{2}\de^{2}_{\xi}f-(q+b\xi)f+|f|^{2}f-\tau_{R}f\de_{\xi}(|f|^2)=0,} where $q$ is the wavenumber of the most intense soliton, proportional to its peak power. Note that in \cite{akhmediev,tran1} the equivalent form of Eq. (\ref{Gagnon}) does not contain the first term $i\de_{x}f$ because in these works only the profile of the stationary solution was considered. To this aim, one can safely ignore the space derivative $\de_{x}f$ to obtain an ODE as done in \cite{akhmediev,tran1}. However, in this work, we also want to investigate the dynamics of the FRS in the accelerated frame ($x,\xi$), and thus we wish to keep the space derivative $\de_{x}f$ in Eq. (\ref{Gagnon}). If one neglects all the nonlinear terms, Eq. (\ref{Gagnon}) would correspond to the classical Schroedinger equation for a unitary mass particle of energy $q$ subject to a 'gravitational' potential $U(\xi) = b\xi$ \cite{gorbach,gorbachnature}.

Equation (\ref{Gagnon}) is non-integrable, and its localised solutions cannot be analytically obtained. Moreover, the soliton is not stationary and the traditional version of the Vakhitov-Kolokolov criterion \cite{vakhitov} and the Evans function approach \cite{Portuguese_papers} seem not to lead to the right conclusions on the stability.

\begin{figure}[htb]
  \centering \includegraphics[width=4in]{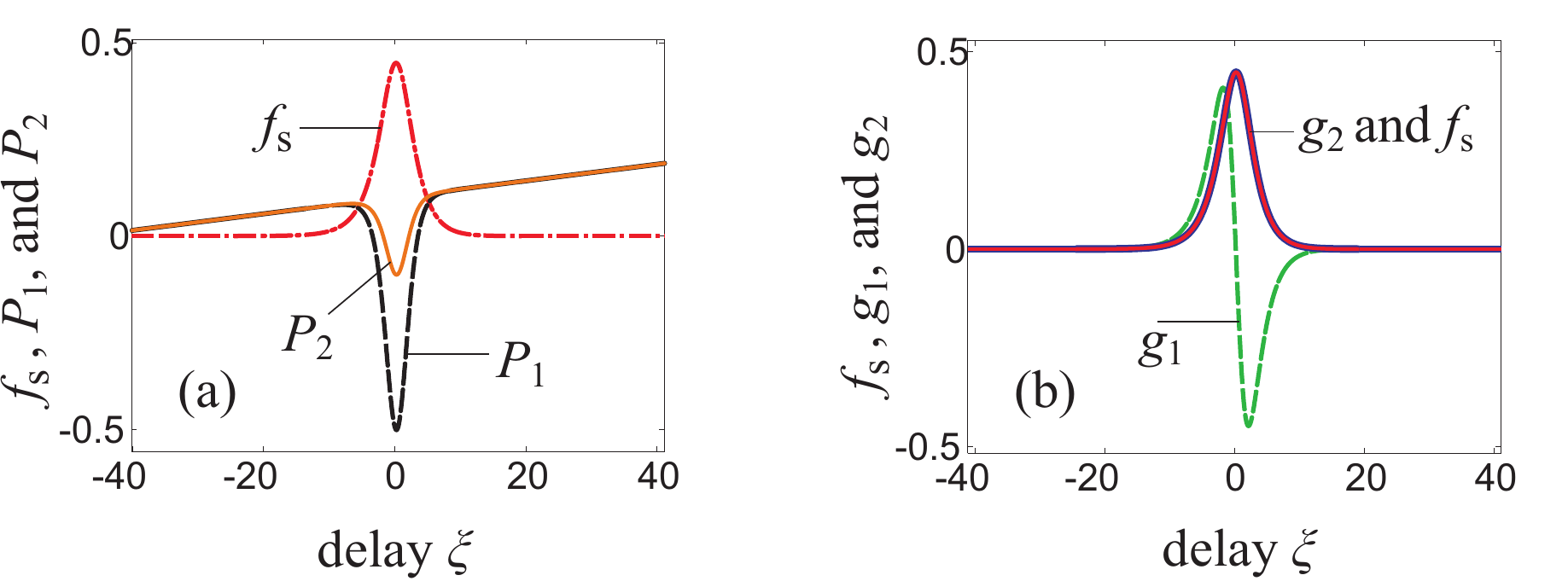}
\caption{\small{(Color online) (a) The profile of the fundamental Raman soliton $f_{\rm s}$ with $q$ = 0.1 (red dashed-dotted curve) and the two potentials $P_{1,2}$(black dashed and orange solid curves, respectively) as functions of the delay $\xi$. (b) The profiles of the two noise eigenfunctions $g_{1}$ (green dashed curve) and $g_{2}$ (red solid curve) corresponding to the two potentials $P_{1,2}$ in (a), respectively.  $g_{1,2}$  are normalised such that the maximum values of $|g_{1,2}|$ are equal to the maximum value of the fundamental Raman soliton $f_{\rm s}$. Note that the normalised $g_{2}$ is identical to $f_{\rm s}$. Raman parameter is $\tau_{R}$ = 0.1.}}
  \label{fig1}
\end{figure}

The profile of the stationary localised solutions to Eq. (\ref{Gagnon}) was numerically obtained in \cite{akhmediev,tran1}. These localised solutions, theoretically, do not change when the space variable $x$ evolves, i.e., $\de_{x}f = 0$. However, if these solutions are not stable, their profiles can still change significantly during propagation as shown later. An example of the FRS profile as a function of the delay $\xi$ is shown in Fig. \ref{fig1}(a) (red dashed-dotted curve labeled $f_{\rm s}$), for a  soliton wavenumber $q = 0.1$. This FRS shown in Fig. \ref{fig1}(a) is a real function, and in what follows we will always assume that $f_{\rm s}$ is real without losing generality. This is due to the properties of Eq. (\ref{Gagnon}) in which if $f_{\rm s}$ is a solution then $f_{\rm s}$exp$(ia)$ is also a solution with $a$ being an arbitrary constant phase. For any value of the soliton wavenumber $q$, such solutions always show a characteristic Airy tail with oscillations on the leading edge of the pulse due to tunnelling in the linearized Schroedinger equation. The Airy tail is quite small when using the physically relevant parameters, and thus, in Fig. \ref{fig1}(a) one cannot see it without strongly zooming the leading edge. Airy tails will be more pronounced if the slope $b$ of the gravity-like potential gets larger, due to the exponential increase of the tunneling probability.

Now we use the linear stability approach to analyse the stability of the FRS. To this end we insert the {\em Ansatz} $f = f_{\rm s} + g$ into Eq. (\ref{Gagnon}), where the complex field $g$ is a weak noise added to the FRS $f_{\rm s}$. After linearising Eq. (\ref{Gagnon}) with respect to $g$ and splitting $g$ into real and imaginary parts ($g = g_{1} + ig_{2}$) one gets the following system of linear equations for the two real fields $g_{1}$ and $g_{2}$:

\begin{eqnarray}
-\de_{x} g_{2}+ \frac{1}{2}\de^{2}_{\xi}g_{1} - (q+b \xi)g_{1} + 3g_{1}f_{\rm s}^{2} - \tau_{R}g_{1}\de_{\xi}(f_{\rm s}^2) - 2\tau_{R}f_{\rm s}\de_{\xi}(f_{\rm s}g_{1}) =0, \label{eigen1}\\
\de_{x}g_{1}+ \frac{1}{2}\de^{2}_{\xi}g_{2} - (q+b \xi)g_{2} + g_{2}f_{\rm s}^{2} - \tau_{R}g_{2}\de_{\xi}(f_{\rm s}^{2}) =0. \label{eigen2}
\end{eqnarray}

It is clear from Eqs. (\ref{eigen1}) and (\ref{eigen2}) that $g_{1}$ and $g_{2}$ are coupled during propagation. However, we first study the stationary solutions of $g_{1,2}$, so one can ignore the terms $\de_{x} g_{2}$ and $\de_{x} g_{1}$ in Eqs. (\ref{eigen1}) and (\ref{eigen2}). In this case, one obtains two following independent linear equations for the noise eigenfunctions:
\begin{eqnarray}
-\frac{1}{2}\de^{2}_{\xi}g_{1} + P_{1}g_{1} + 2\tau_{R}f_{\rm s}^{2}\de_{\xi}g_{1}=0, \label{eigenP1} \\
-\frac{1}{2}\de^{2}_{\xi}g_{2} + P_{2}g_{2} =0,  \label{eigenP2}
\end{eqnarray}
where the potentials felt by $g_{1,2}$ are respectively $P_{1} = q+b\xi -3f_{\rm s}^{2} + 4\tau_{R}f_{\rm s}\de_{\xi}f_{\rm s}$ and $P_{2} = q+b\xi -f_{\rm s}^{2} + 2\tau_{R}f_{\rm s}\de_{\xi}f_{\rm s}$. Note that Eqs. (\ref{eigenP1}) and (\ref{eigenP2}) are formally identical to two stationary Schroedinger equations with potentials $P_{1,2}$, but the third term in Eq. (\ref{eigenP1}) provides {\em gain} for $g_{1}$, leading to a growth for the real part of the noise -- and thus also to the imaginary part $g_{2}$ via the dynamical coupling along the propagation. This will lead to instabilities of the FRS, and is the main result of this Letter.

In Fig. \ref{fig1}(a) we plot $P_{1,2}$ (black dashed and orange solid curves, respectively) as functions of the delay $\xi$ for the specific FRS $f_{\rm s}$.

Now we find the profiles of the two noise eigenfunctions $g_{1,2}$. By numerically solving Eqs. (\ref{eigenP1}) and (\ref{eigenP2}) with the given $f_{\rm s}$, we are able to obtain the localised solutions for $g_{1,2}$ as shown in Fig. \ref{fig1}(b) (green dashed  curve and red solid curve, respectively). For comparison we also plot the FRS $f_{\rm s}$ in Fig. \ref{fig1}(b) (the blue curve hidden behind the red solid curve). As mentioned above, the noise eigenfunctions $g_{1,2}$ must be much weaker than the FRS $f_{\rm s}$. Just for the sake of comparison, the eigenfunctions $g_{1,2}$ plotted in Fig. \ref{fig1}(b) are normalised such that the maximum values of $|g_{1,2}|$ are equal to the maximum value of the FRS $f_{\rm s}$. In Fig. \ref{fig1}(b) one can easily see that the functional shape of $g_{2}$ is identical to $f_{\rm s}$, therefore, the red and blue curves representing these two functions are on top of each other in Fig. \ref{fig1}(b). This is not surprising because if one replaces $g_{2}$ in Eq. (\ref{eigenP2}) by $f_{\rm s}$, then one gets immediately the stationary Eq. (\ref{Gagnon}) (for $i\de_{x}f=0$).

\begin{figure}[htb]
  \centering \includegraphics[width=4in]{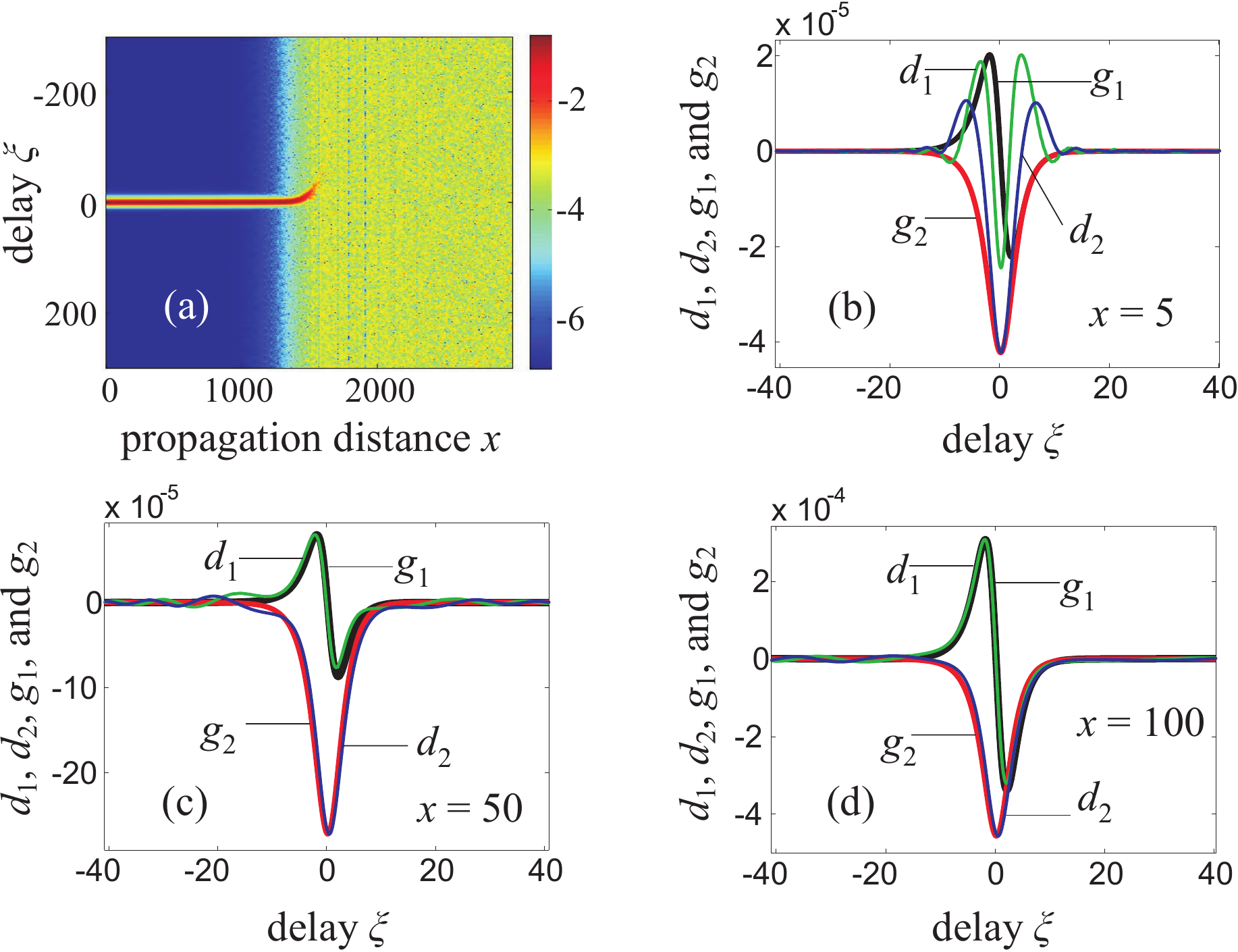}
   \caption{\small{(Color online) (a) Propagation of the fundamental Raman soliton and its instability. (b,c,d) The profiles of the real ($d_{1}$ - green curve) and imaginary parts of the deviation ($d_{2}$ - blue curve) of the FRS profile during propagation as compared to the FRS profile $f_{\rm s}$ at the input for three values of the propagation distance $x$ = 5, 50, and 100, respectively. The profiles of re-normalised eigenfunctions $g_{1,2}$ are also plotted in (b,c,d). Fig. 2(a) is on a logarithmic scale}}
  \label{fig2}
\end{figure}

In what follows we will compare the noise eigenfunctions $g_{1,2}$ (which are theoretically obtained above) with the noise generated during propagation of the FRS. Figure \ref{fig2}(a) shows the propagation of the FRS [whose initial profile is $f_{\rm s}$ as plotted in Fig. \ref{fig1}(a)] in the accelerated reference frame ($x$,$\xi$). As clearly seen in Fig. \ref{fig2}(a), when the propagation distance $x<1000$ the profile of the FRS does not change significantly during propagation. However, for larger propagation distances, the noise grows noticeably, then at a distance $x\simeq1500$ the FRS is completely destroyed by a powerful instability and one only gets an approximately uniform noisy background. We want to emphasise that this behaviour of the FRS during propagation is not an artefact due to the numerical inaccuracy accumulated during the propagation, because one always gets this instability even when the numerical accuracy is greatly enhanced (larger delay window [$\xi_{min}$,$\xi_{max}$], smaller discretisation step for both the $\xi$- and $x$-axis). So, Fig. \ref{fig2}(a) is a good evidence that, in the reduced NLS, {\em the FRS is metastable} like all other Raman multipeak states -- it can propagate for long distances without significantly changing its profile, but finally it will collapse for a specific critical distance. 

Now we define the deviation of the FRS profile at the current value of the propagation distance $x$ from the initial FRS $f_{\rm s}$ profile during its propagation as follows:
\eq{deviation}{d(x,\xi) \equiv d_{1}(x,\xi) + id_{2}(x,\xi) \equiv f(x,\xi)-f_{\rm s}(\xi),}
where $f(x,\xi)$ is the profile of the FRS at the distance $x$ that one gets by numerically solving Eq. (\ref{Gagnon}), with the input field being $f_{\rm s}$, whereas $d_{1}$ and $d_{2}$ are the real and imaginary parts of $d$, respectively. These deviation components $d_{1,2}$ can be interpreted as the real and imaginary parts of the noise during propagation of the FRS. In a stable situation, as mentioned above, this deviation must be equal to zero, i.e., $d(x,\xi)$ = $d_{1}(x,\xi)$ = $d_{2}(x,\xi)$ = 0. However, since the FRS is not stable, this deviation will differ from zero and grow during propagation. In Figs. \ref{fig2}(b,c,d) we plot the real ($d_{1}$ - green curve) and imaginary ($d_{2}$ - blue curve) parts of $d(x,\xi)$ for three values of the propagation distance, $x$ = 5, 50, and 100, respectively. The profiles of the noise eigenfunctions $g_{1,2}$ are also plotted in Figs. \ref{fig2}(b,c,d) with black and red curves, respectively. $g_{1,2}$ in Figs. \ref{fig2}(b,c,d) are re-normalised such that max($g_{1}$) = max($d_{1}$), and min($g_{2}$) = min($d_{2}$). At the beginning of the propagation when $x=5$ [Fig. \ref{fig2}(b)], the theoretically obtained eigenfunctions $g_{1,2}$ differ significantly from the corresponding deviation components $d_{1,2}$. However, for larger propagation distances, as clearly shown in Fig. \ref{fig2}(c) ($x=50$) and Fig. \ref{fig2}(d) ($x=100$), the deviation components of the noise $d_{1,2}$ have profiles which are almost identical to $g_{1,2}$, fully confirming our theoretical analysis based on Eqs. (\ref{eigen1}-\ref{eigen2}).
Note that when the propagation distance is increased, the noise gradually grows in strength, for instance, when $x=5$ [Fig. \ref{fig2}(b)] max($d_{1}$) $\simeq$ $2.10^{-5}$, whereas when $x=100$ [Fig. \ref{fig2}(d)] max($d_{1}$) $\simeq$ $3.10^{-4}$. The profile difference between $d_{1,2}$ and $g_{1,2}$ in Fig. \ref{fig2}(b) is understandable, because the deviation components $d_{1,2}$ of the noise during propagation need time to build up and acquire the profiles of their corresponding eigenfunctions $g_{1,2}$. For much larger propagation distances ($x>1000$), as shown in Fig. \ref{fig2}(a), the noise grows significantly, and as a result the condition that the noise must be much weaker than the FRS is not longer satisfied. Evidently, in this case, the procedure for obtaining the $g_{1,2}$ discussed above is not valid. 

\begin{figure}[htb]
  \centering \includegraphics[width=4in]{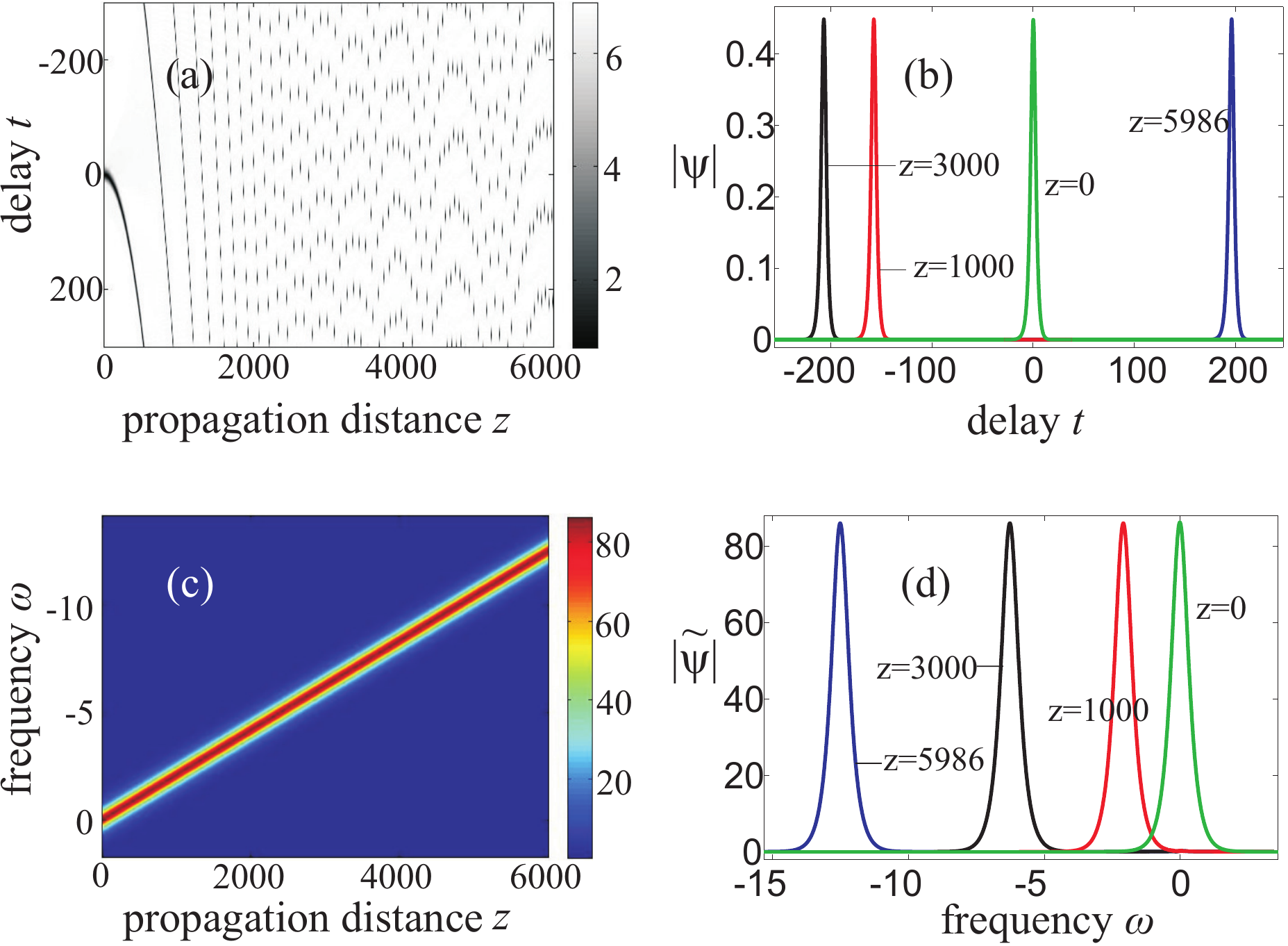}
   \caption{\small{(Color online) (a,c) Temporal and spectral evolution of a FRS using the full Raman convolution, respectively. (b,d) The temporal and spectral profiles of the FRS, respectively, for different propagation distances $z$. }}
  \label{fig3}
\end{figure}

In Fig. \ref{fig3}(a) we show the propagation of a FRS in the laboratory frame $(z,t)$ when using the full Raman convolution, see for instance Eq. (2.3.36) of Ref. \cite{agrawalbook1}. We use 3000 frames along the $z$-axis to show results in Fig. \ref{fig3}(a). After reaching the bottom part of the time window, the soliton will re-emerge in the upper part due to the periodic boundary conditions. However, since there is never self-interaction and the time windows is large, this simulates an undisturbed and unlimited propagation. We have propagated the FRS for 6000 dispersion lengths and concluded that {\em there is no instability} that leads to the soliton collapse. In Fig. \ref{fig3}(c) we show the spectral evolution of the FRS during propagation. The temporal and spectral profiles of the FRS are plotted in Figs. \ref{fig3}(b) and (d), respectively, for different propagation distances. As clearly shown in Fig. \ref{fig3}, both temporal and spectral profiles of the FRS are well conserved during propagation. It is important to note that the input condition for simulation results in Fig. \ref{fig3}(a) is the numerically obtained $f_{s}$ of the reduced NLSE shown in Fig. \ref{fig1}(a), which is not the exact soliton solution of the NLSE with the full Raman convolution. One can thus conclude that the physical FRS is perfectly stable.

In conclusion, we have investigated the stability issue of the fundamental Raman soliton of the reduced nonlinear Schroedinger equation in optical fibres, where the Raman gain is approximated as linear. We demonstrate theoretically that the FRS is metastable in the reduced NLSE, in the sense that it can propagate several hundreds, even thousands of dispersion lengths along the fibre without significant changes in its profile. However, with larger propagation distances the noise is amplified, and finally, the structure of the fundamental Raman soliton completely collapses due to a powerful instability, confirming and explaining previous numerical observations \cite{erkintalo1,solli1}. The eigenfunctions of the noise obtained via our simple theoretical approach agree very well with the noise of the FRS obtained directly through pulse propagation simulations. However, we show that the FRS of the full NLSE is perfectly stable. This underlines the dangers of using approximations of the Raman response functions in any dynamical simulations.

This work is supported by the German Max Planck Society for the Advancement of Science (MPG). F.B. would like to thank M. Erkintalo, J. Dudley and G. Genty for useful comments.


\begin{thebibliography}{99}

\bibitem{agrawalbook1} G.P. Agrawal, {\em Nonlinear fibre Optics}, 5th ed. (Academic Press, 2013).


\bibitem{kivshar} Y.S Kivshar and G.P. Agrawal, {\em Optical Solitons: from fibre to Photonic  Crystals}, 5th ed. (Academic, 2003).

\bibitem{akhmediev} N. Akhmediev, W. Kr\'olikovski, and A. J. Lowery, ``Influence of the Raman-effect on solitons in optical fibers," Opt. Comm. {\bf 131}, 260-266 (1996).

\bibitem{pod1} A. Podlipensky, P. Szarniak, N. Y. Joly, C. G. Poulton, and P. St. J. Russell, ``Bound soliton pairs in photonic crystal fiber," Opt. Express {\bf 15}, 1653-1662 (2007).

\bibitem{pod2} A. Podlipensky, P. Szarniak, N. Y. Joly, and P. St. J. Russell, ``Anomalous pulse breakup in small-core photonic crystal fibers," J. Opt. Soc. Am. B {\bf 25}, 2049-2056 (2008).

\bibitem{russell} P. St. J. Russell, ``Photonic crystal fibers," Science {\bf 299}, 358-362 (2003).

\bibitem{tran1} Tr. X. Tran, A. Podlipensky, P. St. J. Russell, and F. Biancalana, ``Theory of Raman multipeak states in solid-core photonic crystal fibers," J. Opt. Soc. Am. B {\bf 27}, 1785-1791 (2010).

\bibitem{tran2} A. Hause, Tr. X. Tran, F. Biancalana, A. Podlipensky, P. St. J. Russell and F. Mitschke, ``Understanding Raman-shifting multipeak states in photonic crystal fibers: two convergent approaches," Opt. Lett. {\bf 35}, 2167-2169 (2010).

\bibitem{Portuguese_papers} M. Facao, {\em Existence and stability of accelerating optical solitons}, Ph.D. thesis, Univ. of Edinburgh (2003).

\bibitem{erkintalo1} M. Erkintalo, G. Genty, B. Wetzel and J. Dudley, ``Limitations of the linear Raman gain approximation in modeling broadband nonlinear propagation in optical fibers," Opt. Express {\bf 18}, 25449-25460 (2010).

\bibitem{solli1} D. R. Solli, C. Ropers, P. Koonath and B. Jalali, ``Optical rogue waves," Nature {\bf 450}, 1054-1057 (2007).

\bibitem{mitschke} F. M. Mitschke and L. F. Mollenauer, ``Discovery of the soliton self-frequency shift," Opt. Lett. {\bf 11}, 659-661 (1986).

\bibitem{gordon} J. P. Gordon, ``Theory of the soliton self-frequency shift," Opt. Lett. {\bf 11}, 662-664 (1986).


\bibitem{belanger} L. Gagnon and P. A. B\'elanger, ``Soliton self-frequency shift versus Galilean-like symmetry," Opt. Lett. {\bf 15}, 466-468 (1990).

\bibitem{gorbach} A. V. Gorbach and D. V. Skryabin, ``Theory of radiation trapping by the accelerating solitons in optical fibers," Phys. Rev. A {\bf 76}, 053803 (2007).

\bibitem{gorbachnature} A. V. Gorbach and D. V. Skryabin, ``Light trapping in gravity-like potentials and expansion of supercontinuum spectra in photonic-crystal fibres," Nature Photon. {\bf 1}, 653-657 (2007).


\bibitem{vakhitov} N. G. Vakhitov and A. A. Kolokolov, ``Stationary solutions of the wave equation in the medium with nonlinearity saturation," Radiophys. Quantum Electron. {\bf 16}, 783 (1973).


\end{thebibliography}
\end{document}